\documentclass[10pt]{article}
\usepackage[OE]{express}
\usepackage{placeins}
\usepackage[utf8]{inputenc}    
\usepackage{siunitx}		
\usepackage{graphicx}
\usepackage{subcaption}

\begin{document}
\title{Ptychographic characterisation of polymer compound refractive lenses manufactured by additive technology}
\author{Mikhail Lyubomirskiy,\authormark{1,*} Frieder Koch,\authormark{2}  Ksenia A. Abrashitova,\authormark{3,4} Vladimir O. Bessonov,\authormark{3,5} Natalia Kokareva,\authormark{3} Alexander Petrov,\authormark{3,4} Frank Seiboth,\authormark{1}  Felix Wittwer,\authormark{1,6} Maik Kahnt,\authormark{1,6} Martin Seyrich,\authormark{1,6} Andrey A. Fedyanin,\authormark{3} Christian David,\authormark{2} and Christian G. Schroer\authormark{1,6}}
\address{
\authormark{1}Deutsches Elektronen-Synchrotron DESY, Notkestra{\ss}e 85, 22607 Hamburg, Germany\\
\authormark{2}Laboratory for Micro- and Nanotechnology, Paul Scherrer Institute, 5232 Villigen-PSI, Switzerland\\
\authormark{3}Faculty of Physics, Lomonosov Moscow State University, Moscow 119991, Russia\\
\authormark{4}Immanuel Kant Baltic Federal University, 14 Nevskogo, 236041 Kaliningrad, Russia\\
\authormark{5}Frumkin Institute of Physical Chemistry and Electrochemistry, Russian Academy of Sciences, Moscow 119071, Russia\\
\authormark{6}Department Physik, Universit{\"a}t Hamburg, Luruper Chaussee 149, 22761 Hamburg, Germany
}

\email{\authormark{*}mikhail.lyubomirskiy@desy.de}

\newcommand{\Fig}[1]{Fig.~\ref{#1}}
\newcommand{\equ}[1]{Eq.~\ref{#1}}

\begin{abstract}

The recent success in the development of high precision printing techniques allows one to manufacture free-standing polymer structures of high quality.
Two-photon polymerization lithography is a mask-less technique with down to $100\,\si{\nano\meter}$ resolution that provides full geometric freedom.
It has recently been applied to the nanofabrication of X-ray compound refractive lenses (CRLs).
In this article we report on the characterization of two sets of CRLs of different design produced by two-photon polymerization induced lithography.
\end{abstract}

\ocis{(340.0340) X-ray optics; (220.3630) Lenses; (220.3740) Lithography.}

\section{Introduction}

X-ray lenses are relatively young optical devices -- the very first experiment on focusing of X-rays by refractive lens was reported only two decades ago \cite{SKSL1996}.
Since then, X-ray lenses experienced a rapid progress and are now widely used at synchrotron radiation sources around the world \cite{LS2016}.
They are made of many different materials, such as Be, Al, Si, SU-8, or Diamond.
The most popular lens type is the doubly-curved rotationally parabolic Be lens \cite{SKLGKBRSSS2002}.
They combine a large physical aperture with a high transmission and can deliver a high flux to the sample.
Their best reported resolution is in the order of \SI{100}{\nano\meter} \cite{SHPSU2013,SSSWRWUNRPGVWRBGFGLNS2017}.
Despite the high flux, these lenses are not very popular for nanoprobe applications due to their limited resolution and the presence of aberrations \cite{SSSWRWUNRPGVWRBGFGLNS2017}.

Higher spatial resolutions have been achieved with Si planar lenses, so-called nanofocusing lenses (NFLs), which are manufactured by planar structuring of silicon.
Focal spot sizes on the order of \SI{50}{\nano\meter} and below are achieved routinely.
NFLs are used in X-ray scanning and even full-field microscopes at PETRA~III and ESRF \cite{SKPBFLBRVHK2006,SBFPSSSSFWR2009,SSLJHDP2016}.
The record resolution of \SI{18}{\nano\metre} was achieved with adiabatically focusing lenses \cite{SL2005c,PKHRSWJRWBSSBFS2017}.
Despite the high resolution Si NFLs have a drawback: their relatively high absorption is limiting the total flux in the nanobeam, moreover the technologically limited depth of etching narrows the physical aperture of the lens, which reduces the total flux in the focus due to the crossed geometry.

Two-photon polymerization induced lithography has attracted increasing attention in particular due to the possibility to manufacture lenses out of amorphous materials with very high three-dimensional shape accuracy.
Several groups have already reported X-ray compound refractive lenses (CRLs) manufactured by two-photon absorption lithography \cite{PBAKSBEKLYPSFS2017, app8050737}.
So far, the quality and performance of those lenses have not assessed experimentally.
In this article we report on the ptychographic characterization of two lens sets, one stacked vertically like a tower, the other stacked horizontally like a train, manufactured by two-photon absorption lithography.

\section{Compound refractive lens (CRL) manufacturing}

Two-photon absorption lithography is based on nonlinear absorption in the focal volume.
The threshold character of the polymerization process provides \SI{100}{\nano\meter} scale resolution, while the nonlinear character of absorption ensures transparency of out-of-focus material and controllable intensity within focal volume.
To provide sufficient intensity values (above the threshold) ultrashort IR lasers and high-numerical-aperture objectives are used.
The materials that are implemented for this lithographic technique include amorphous epoxy resists (SU-8), acrylate resists (IP series from Nanoscribe), and hybrid organic-inorganic resists (ORMOCOMP, SZ 2080).

\begin{figure}[!htbp]
\centering\includegraphics[width=6.5cm]{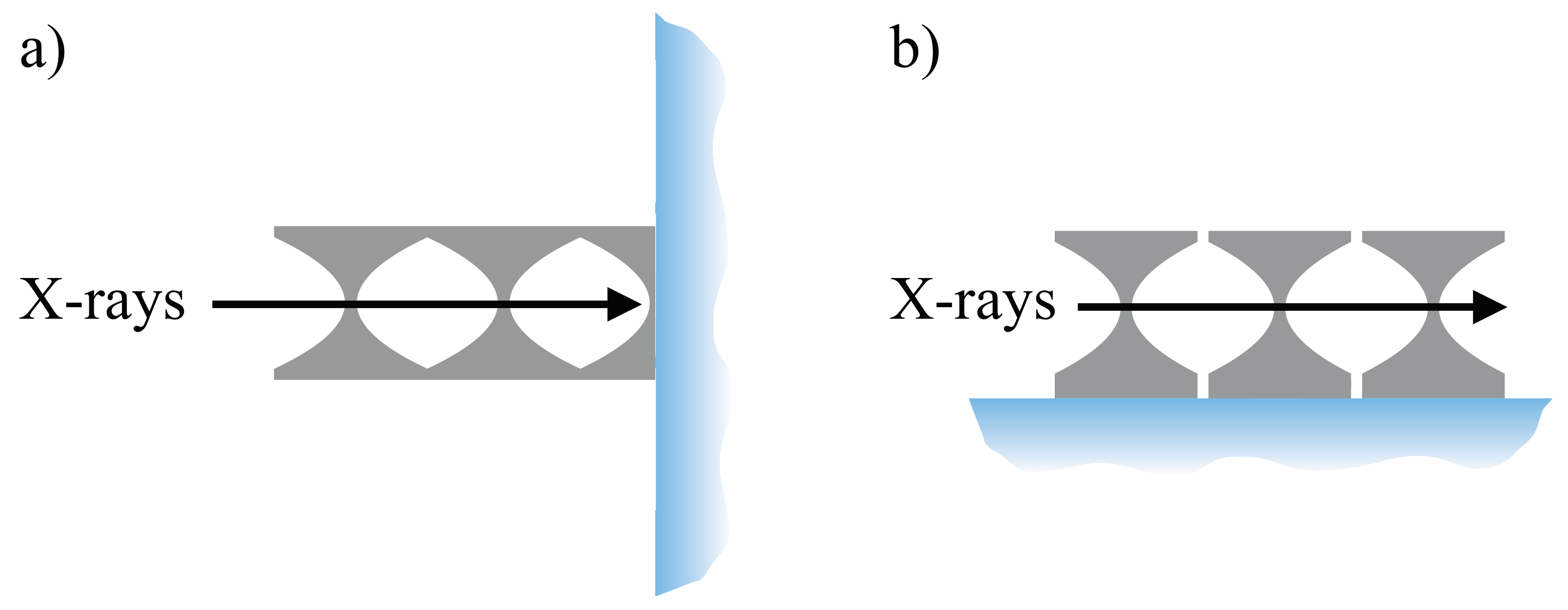}
\caption{Compound refractive lens geometries fabricated by additive machining: (a) vertical or tower, (b) horizontal or train geometry.}
\label{lens_geometry}
\end{figure}

In general, there are two geometries to print CRLs on a substrate: as a lens tower where the optical axis is perpendicular to the substrate (\Fig{lens_geometry} (a) -- vertical orientation) and as a lens train where the optical axis is parallel to the substrate (\Fig{lens_geometry} (b) -- horizontal orientation).
We have manufactured and tested lenses of both types.

\FloatBarrier

\subsection{Compound refractive lenses in vertical or tower design}

In the vertical geometry with the optical axis perpendicular to the substrate [cf.\ Fig.~\ref{lens_geometry}(a)], the optical axis of the lens can be aligned with that of the laser exposure system.
This is advantageous because the printing voxel is anisotropic, with the extent along the optical axis about $2.5$ times the lateral extent, depending on the objective \cite{FW2013}.
Consequently, in this configuration one can write the curvatures in the two focusing directions with the same resolution and local crosslinking strain characteristics.
This is expected to lead to minimized astigmatism.

The lens structures with this geometry were fabricated using a Nanoscribe \textit{Photonic~Professional~GT} at the Laboratory of Micro- and Nanotechnology of the Paul Scherrer Institute (PSI).
We used the Nanoscribe IP-S resist in dip-in lithography mode, with a $25\times$ objective ($\mbox{NA}=0.8$).
The lateral voxel displacement was effectuated by the system's mirror galvanometer system, providing maximum writing speed, while a piezo motor was used for displacement along the optical axis.
The piezo travel range sets a limit of \SI{300}{\micro\meter} on the length of the CRL, which could be extended to the \si{\milli\meter} range with stitching in $z$ direction using the mechanical microscope stage.
Both hatching and slicing of the lens geometry for exposure were performed with a spacing of only \SI{100}{\nano\meter}, which is significantly less than the voxel size, to minimize surface roughness.
Furthermore, the vertical geometry requires a high X-ray transmission substrate, in this case a \SI{250}{\nano\meter} thick $\text{Si}_{\text{3}} \text{N}_{\text{4}}$ membrane.

\begin{figure}[!htbp]
\centering\includegraphics[width=10cm]{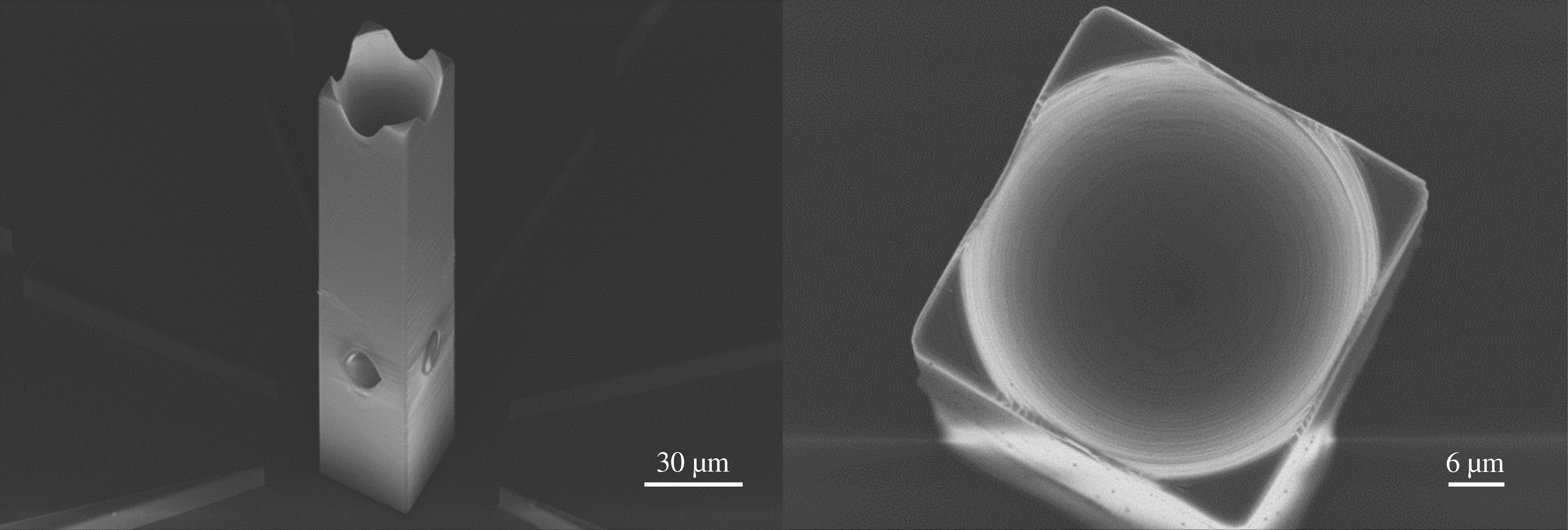}
\caption{Scanning electron microscopy image of compound refractive lens (left) in vertical or tower geometry and of its entrance aperture (right).}
\label{PSI_SEM}
\end{figure}

To estimate the refractive index decrement, we use the chemical composition of the main constituent, the acrylate resin ($\text{C}_{\text{14}} \text{H}_{\text{18}} \text{O}_{\text{7}}$), and a density of \SI{1.2}{\gram\per\cubic\centi\metre}, which is \SI{2}{\percent} higher than the uncrosslinked resin density.
Using these parameters, we obtain a refractive index decrement of \num{3.94e-6} at \SI{8.2}{\kilo\electronvolt} photon energy.
The parabola apex radius $R$ was chosen as \SI{3}{\micro\meter} and the aperture of the lens was a square with \SI{42}{\micro\meter} edge length.
This geometry allows three curved surfaces to be printed within the range of the piezo scanner without making use of stitching with the mechanical microscope stage; see \Fig{PSI_SEM} for scanning electron microscope images of the resulting structure.
The rotational paraboloid of the lens has to be cut on at least one side to allow the developer liquid access to the whole geometry, resulting in a square aperture with rounded edges.
\FloatBarrier

\subsection{Compound refractive lenses in horizontal or train design}
\begin{figure}[!htbp]
\centering
\includegraphics[width=10cm]{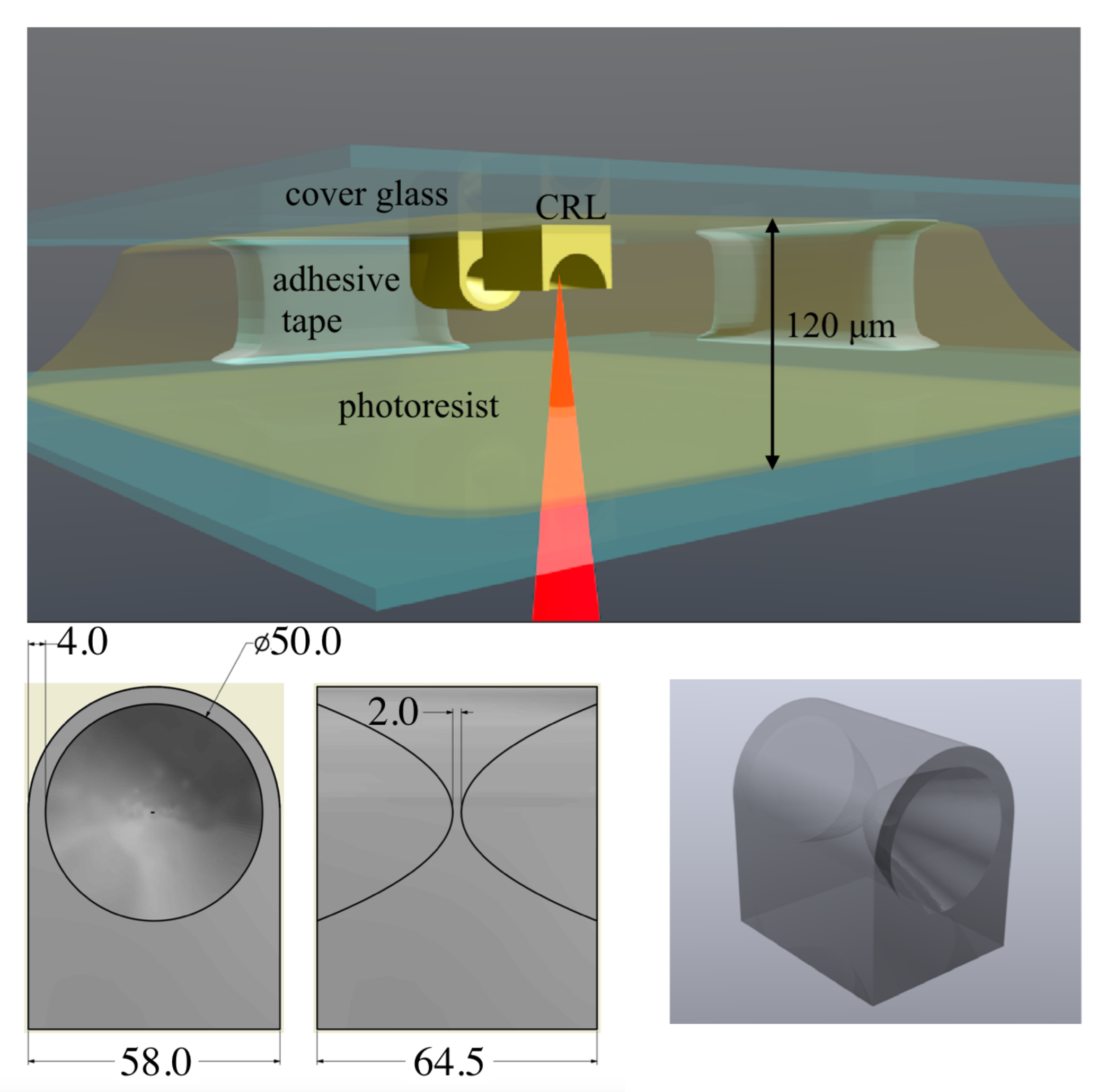}
\caption{Schematic representation of the printing process in case of horizontally organized CRL (top) and geometrical parameters of an individual lens in horizontal design (bottom-left) and 3D model of an individual lens (bottom-right).
All distances are in \SI{}{\micro\meter}.}
\label{model}
\end{figure}

For this design, the ORMOCOMP photoresist from Micro Resist Technology GmbH was used for microfabrication.
The molecular formula is  $\text{C}_{\text{21}} \text{S} \text{O}_{\text{8}}  \text{Si} \text{H}_{\text{36}}$.
The density of cross linked polymer material is estimated to be \SI{1.22}{\gram\per\cubic\centi\metre}.
The exposure of the resist was performed from top to bottom in a cell containing uncrosslinked photoresist.
The thickness of the cell was approximately \SI{120}{\micro\meter} and was controlled by two pieces of adhesive tape (\Fig{model}, top).

A fast steering mirror provided the beam-waist movement within the sample plane ($XY$) with the accuracy of \SI{3}{\nano\meter} in the field of \SI{150}{\micro\meter} $\times$ \SI{150}{\micro\meter} and a piezoelectric stage moved the laser beam waist in a perpendicular direction ($Z$) with the accuracy of \SI{5}{\nano\meter} in the travel range of \SI{200}{\micro\meter}.
High-aperture oil-immersion objective ($\mbox{NA} = 1.4$) focused radiation from Vitara Coherent Ti:sapphire femtosecond laser (\SI{800}{\nano\meter}, \SI{80}{\mega\hertz}, \SI{50}{\femto\second}).
The exposure parameters --- average power of incident radiation and beam waist speed --- were \SI{26}{\milli\watt} and \SI{1200}{\micro\metre\per\second}, respectively.

The printing process was conducted in a layer-by-layer manner with linear infill and 3 perimeters.
To provide better mechanical stability and enhanced adhesion to the substrate first $50$ layers were printed with \SI{200}{\nano\meter} slicing and hatching distances.
The remaining layers had the slicing and hatching distances of \SI{250}{\nano\meter}.

Finite values of field of view and of the working distance of focusing objective imposed limitations on CRL geometry.
The aperture of individual lens was  $R_0 = \SI{50}{\micro\metre}$, and the radius of curvature of parabolic profile was $R = \SI{10}{\micro\metre}$ (\Fig{model}, bottom).
Two sets of horizontally stacked CRLs were manufactured.
They comprised of $8$ and $9$ individual lenses printed in a row.
The distance between individual lenses was chosen to be \SI{120}{\micro\meter} in order to minimize the proximity effect.

\begin{figure}[!]
\centering
\includegraphics[width=10cm]{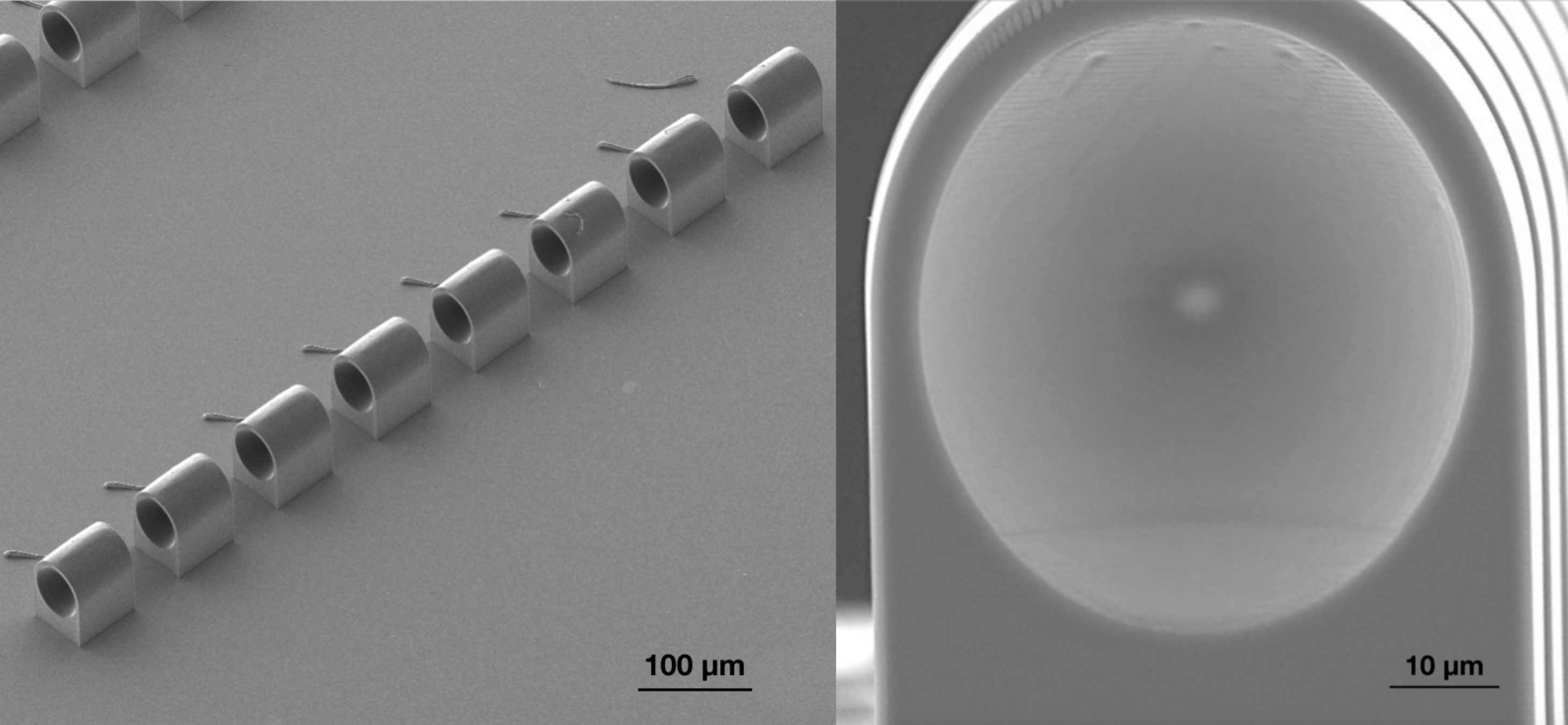}
\caption{Scanning electron microscopy images of the compound refractive lens in horizontal design composed of 8 individual lenses (left) and magnified image of the entrance aperture of an individual lens (right).}
\label{MSU_SEM}
\end{figure}

The CRLs were characterized by scanning electron microscopy (SEM).
The SEM images show that individual lenses suffer from slight ellipticity.
The difference between minor and major axis was approximately \SI{10}{\percent}.
A horizontal defect is also observable in the lower part of the entrance aperture [see \Fig{MSU_SEM}(right)].

\FloatBarrier

\section{X-ray optical characterization of the Lenses}

Consequently even if the object is out of focus, it is possible to propagate the reconstructed wavefield numerically to the focal plane and characterize the focal spot.
To date ptychography has proven to be a robust technique and is a standard technique for optics characterization \cite{SBFHPSSGWSGMVWSBS2010,KTDBMVJP2010,KGLQSBKVBFMA2010,HHPSSSBS2011c,VDGMKMBD2011,SHPSU2013,KBNKPHLS2014,SSPHWRSKJRBFS2014,MPAKMPGBBBOYACB2015,PKHRSWJRWBSSBFS2017}.

\begin{figure}[!htbp]
\centering
\includegraphics[width=13cm]{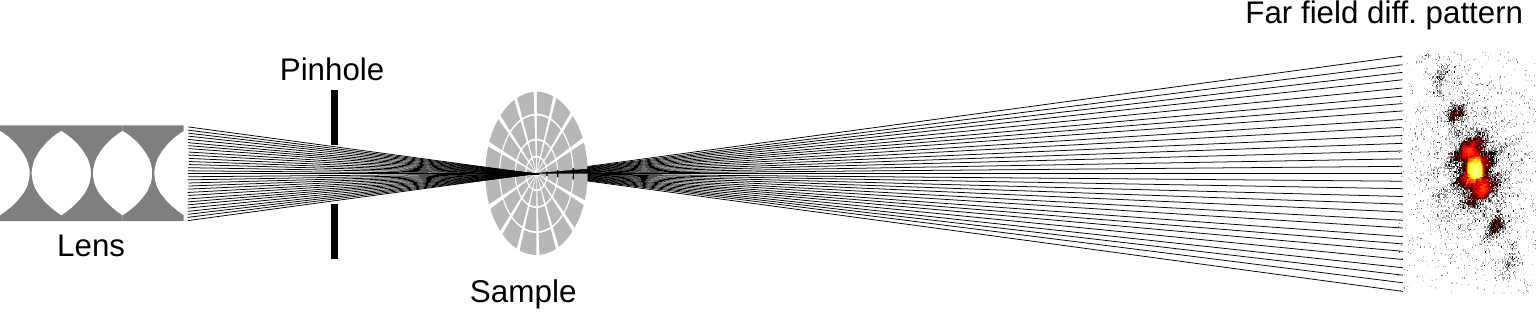}
\caption{Schematic drawing of the ptychographic experimental setup}
\label{scheme}
\end{figure}

The principal experimental scheme is depicted in \Fig{scheme}.
A double crystal Si-(111) monochromator was used to select the desired X-ray energy of \SI{8.2}{\kilo\electronvolt} at I13-1 and \SI{9.0}{\kilo\electronvolt} at P06.
In order to cut-off undesired higher harmonics a pair of total reflection mirrors was used.
Lenses were aligned in the beam using translation and rotation stages.
Downstream of the lenses and close to their focal plane a two-dimensional resolution test chart (Siemens star) manufactured by NTT-AT was placed.
Air scattering downstream of the sample was suppressed by a Helium filled flight tube at I13-1 and an  evacuated tube at P06.
The distance between source and lens position was around \SI{200}{\meter} at I13-1 and around \SI{98}{\meter} at P06.

According to the well known X-ray lens equation \cite{SKSL1996}, a thin compound lens focuses a parallel beam at a distance
\begin{equation}
f = R/(2N\delta),
\end{equation}
where $f$ is the focal length and $N$ is the number of doubly-curved lenses in a stack.
The refractive index in the X-ray range is typically written in the form $n=1-\delta+i\beta$, where $\delta$ is the refractive index decrement (real part) and $\beta$ describes the attenuation inside the material.
The size of the focal spot can be estimated as a convolution of the demagnified geometrical image of the source and the Airy disc, whose width for refractive optics is given by\cite{LSBGGKMZ2002c}
\begin{equation}
d_{\mbox{\footnotesize{diff}}} = 0.75\frac{\lambda}{2 \mbox{NA}},
\label{d_size}
\end{equation}
where $\mbox{NA} = D_{\mbox{\footnotesize{eff}}}/2f$ is the numerical aperture, $D_{\mbox{\footnotesize{eff}}}$ is the effective aperture, and $\lambda$ is the wavelength of the incident radiation \cite{LSBGGKMZ2002c}.

\begin{figure}[!htbp]
\centering
\includegraphics[width=13cm]{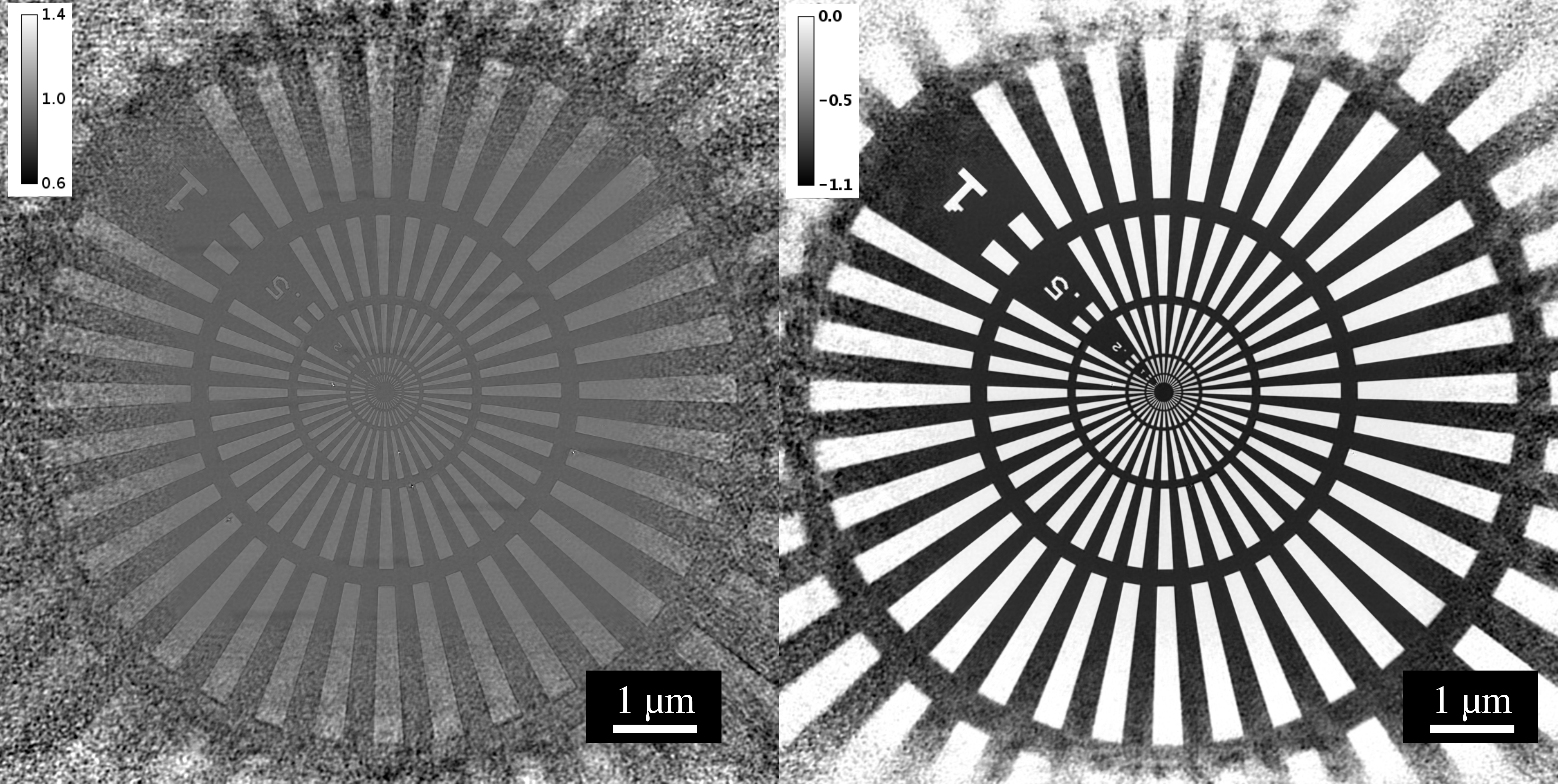}
\caption{Reconstructed amplitude (left) and phase (right) profiles of the NTT-AT test sample}
\label{siemensstar}
\end{figure}

In order to characterize the beam produced by the lenses, the test pattern was scanned in the estimated focal plane.
\Fig{siemensstar} shows an example of the reconstructed phase and amplitude of the test object from one of the scans.
\FloatBarrier

\subsection{Characterization of CRLs in tower design}
\begin{figure}[!htbp]
\centering
\includegraphics[width=10cm]{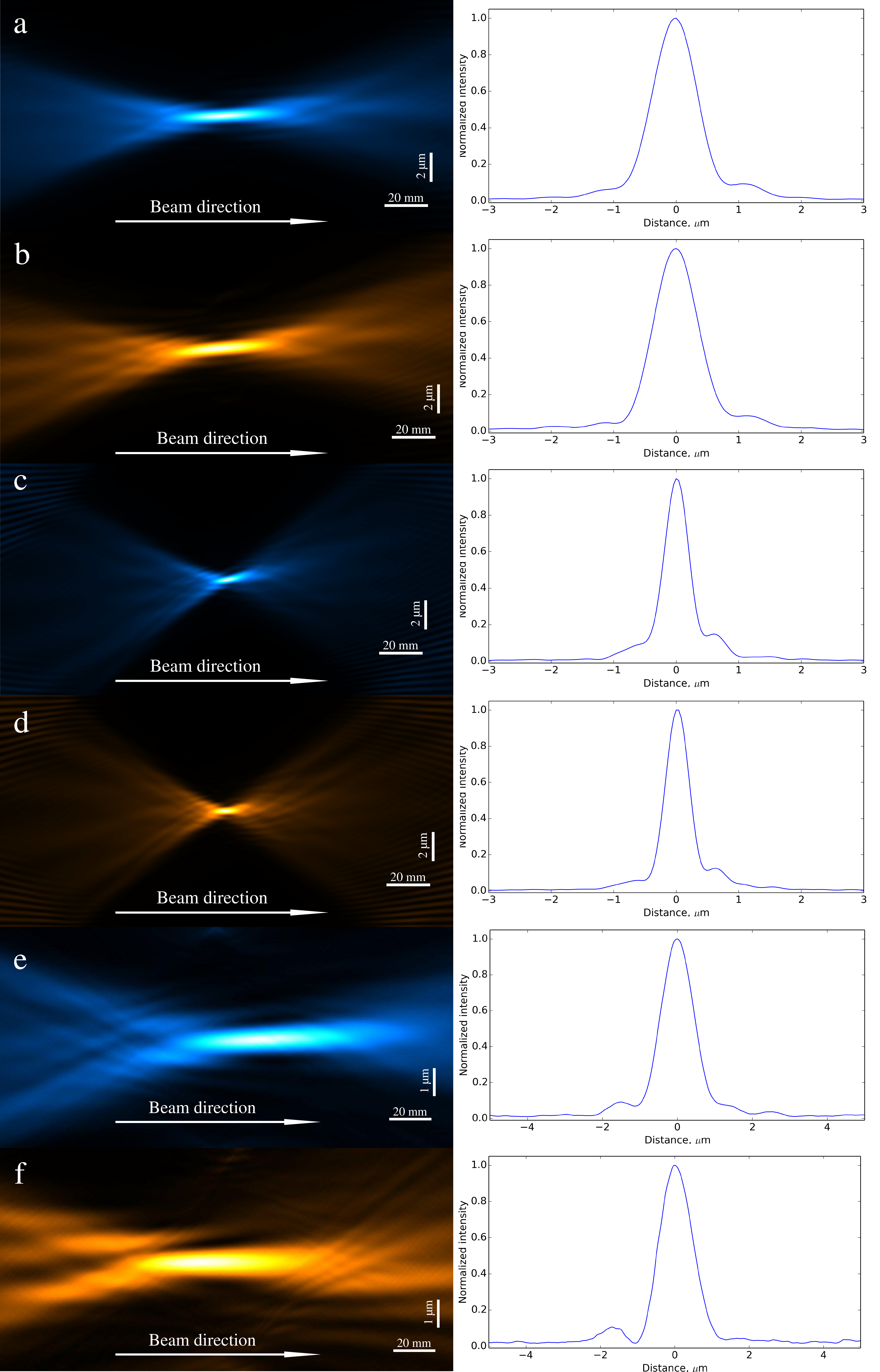}
\caption{Left: Beam profiles, in horizontal (Blue) and vertical (Yellow) directions, generated by numerical propagation of reconstructed wave fields.
Right: Cross sections of the profile through the focus plane.}
\label{all_profiles}
\end{figure}

In the first step we have characterized a single free-standing CRL.
The test pattern was placed at a lens focal distance of $f = \SI{240}{\milli\meter}$ downstream from the lens, which is three orders of magnitude smaller than the distance to the source.
Thus, the image distance of the source approximately coincides with the focal length.
Figures \ref{all_profiles} (a) and (b) show the beam profiles in the horizontal and vertical planes, respectively, together with their cross sections through the focus.
The latter were generated by numerical propagation of the wave field reconstructed via ptychography into the focal plane.
The full-width at half maximum (FWHM) focal spot size is $d = \SI[separate-uncertainty]{820(3)}{\nano\meter}$ as compared to the theoretically expected one of $d =\SI{680}{\nano\meter}$.
The mismatching could be caused by deviations of the lens profile from ideal parabolic shape.

By reducing the focal length, the numerical aperture and be increased, reducing the focal spot size.
This is achieved by combining two CRLs with tower design.
The test sample was placed into the focus at the reduced distance of $f = \SI{120}{\milli\meter}$.
Figures \ref{all_profiles} (c) and (d) contain the beam profiles in horizontal and vertical planes, respectively, and the corresponding cross sections of the beam.
The size of the focal spot (FWHM) is $d = \SI[separate-uncertainty]{490(2)}{\nano\meter}$, which is slightly larger than theoretically expected $d = \SI{420}{\nano\meter}$.
The mismatching could be caused by deviations of the lens profile from ideal parabolic shape.

In addition we have performed a test of the second set of lenses with similar parameters at the beamline P06 at PETRA III.
Figures \ref{all_profiles} (e) and (d) show the beam profiles in the horizontal and vertical plane, respectively, and their cross sections of the beam produced by two doubly-curved lenses with curvature of $R = \SI{2.5}{\micro\metre}$.
The lenses showed a similar performance as the first set.
The size of the focal spot was $d = \SI[separate-uncertainty]{1015(4)}{\nano\meter}$ (FWHM) while the theoretically expected one is $d = \SI{980}{\nano\meter}$.
\begin{figure}[!htbp]
\centering
\includegraphics[width=10cm]{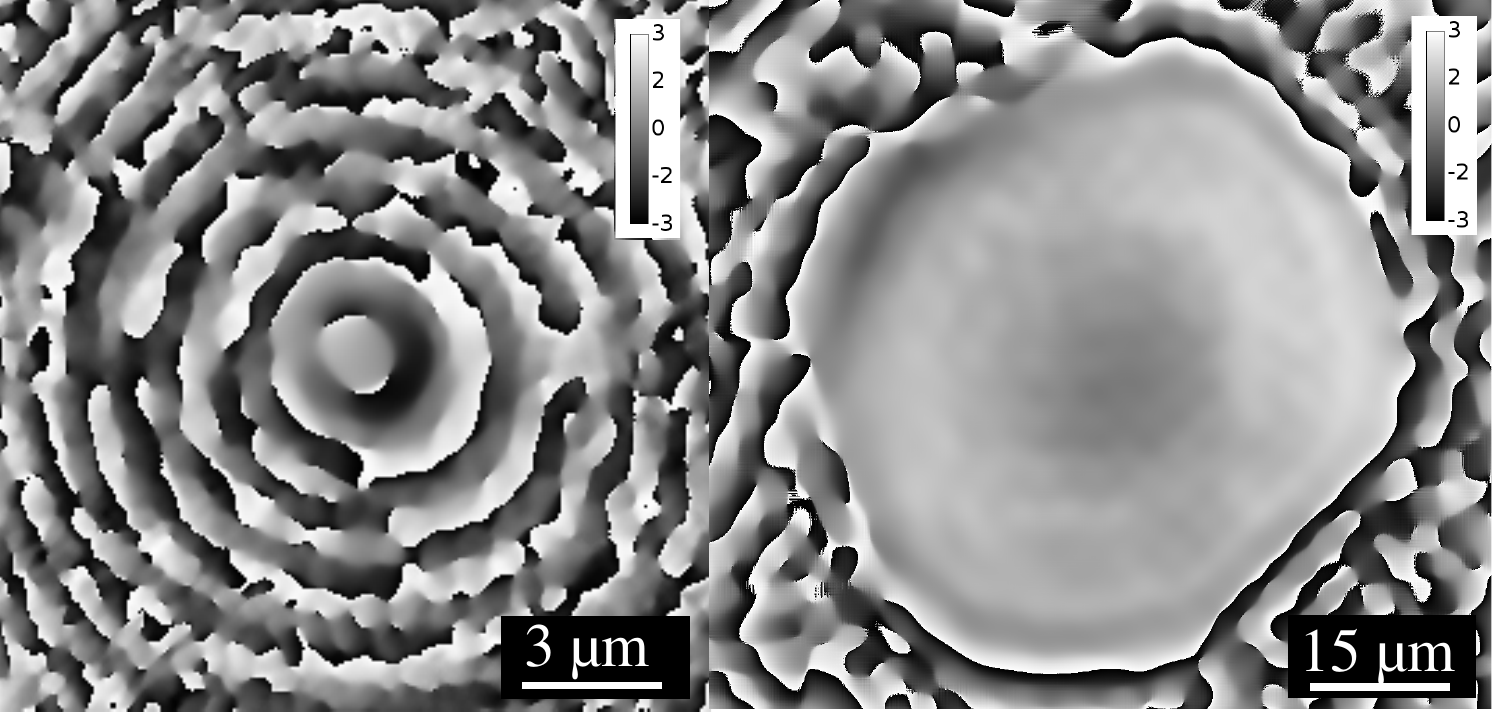}
\caption{Left: Phase profile of the beam at the focal plane. Right: Phase error induced by the combined stack of vertically organized CRLs (phase profile of the beam with subtracted spherical wave)}
\label{2017+2018A_prof}
\end{figure}
\FloatBarrier
The results obtained for vertically arranged lenses showed that they are almost devoid of spherical aberrations.
The phase profile of the focus (\Fig{2017+2018A_prof} (a)) as well as phase profile of the lens exit plane (\Fig{2017+2018A_prof} (b)) is completely symmetrical which confirms the absence of astigmatism of the focal spot.
A slight mismatching of measured focal spot size with calculations could be induced by shrinking of the lens material during developing process.
With knowledge of the degree of shrinkage future lenses can be pre-corrected by stretching of the model.

\subsection{Characterization of CRLs in train design}

Two lens stacks comprised of 9 and 8 individual lenses were studied ($9$-CRL and $8$-CRL respectively).
The theoretical values for the focal lengths of the lenses are $f = \SI{137}{\milli\meter}$ for 9-CRL and $f = \SI{154}{\milli\meter}$ for $8$-CRL.

\begin{figure}[!htbp]
\centering\includegraphics[width=10cm]{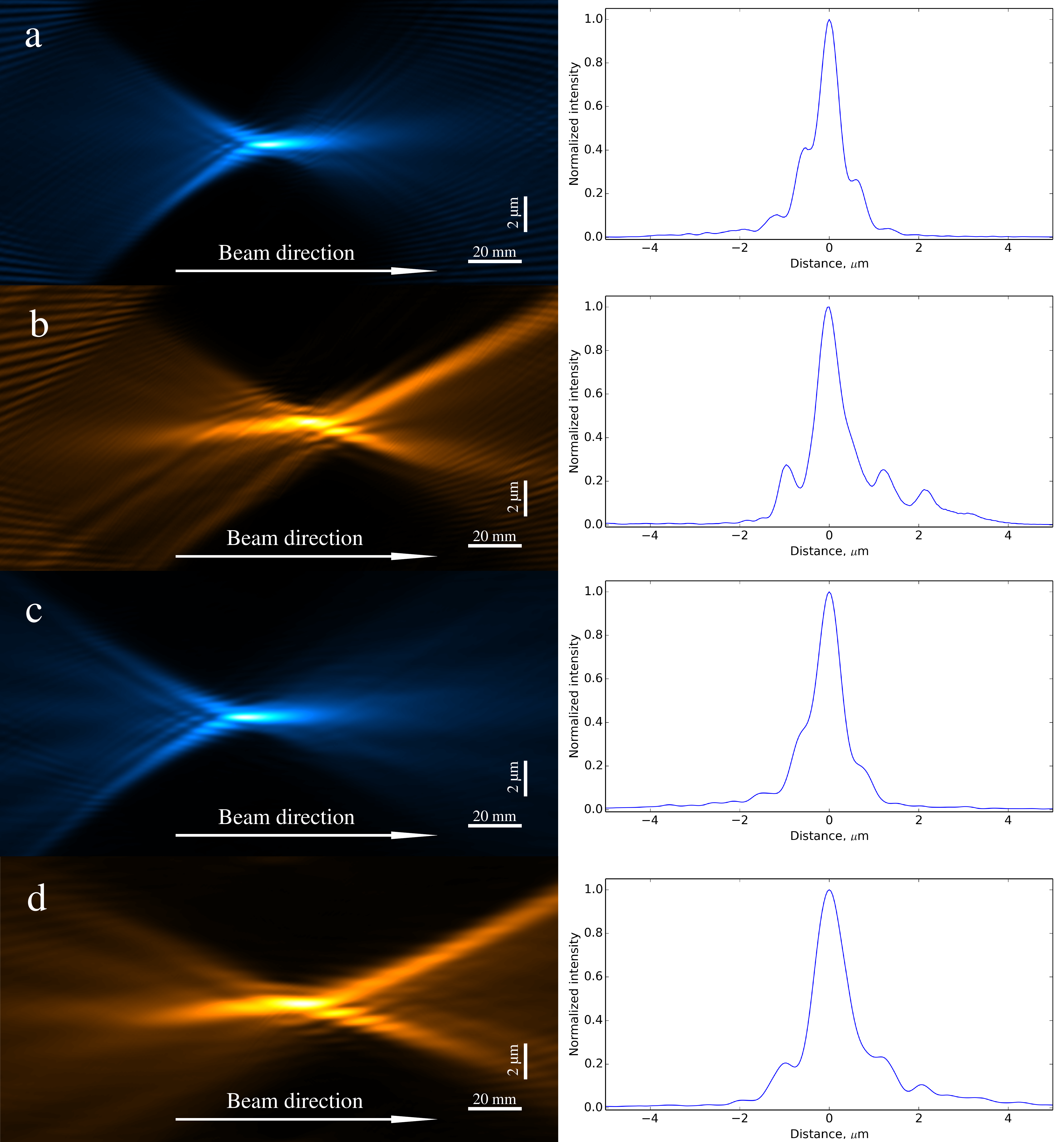}
\caption{Left: Beam profiles, in horizontal (Blue) and vertical (Yellow) directions, generated by numerical propagation of reconstructed wave fields.
Right: Cross sections of the profile through the focus plane.}
\label{9lens_prof}
\end{figure}

Figures \ref{9lens_prof} (a) and (b) show beam profiles in horizontal and vertical directions, respectively, and their cross sections, generated by $9$-CRL.
Side lobes upstream of the focal spot on the horizontal profile and downstream of the focus on the vertical profile indicate the presence of spherical aberrations.
Moreover it is directly confirmed by the mismatching of the focus positions on the vertical and the horizontal beam profiles.
The measured size of the focal spot is $d = \SI[separate-uncertainty]{560(17)}{\nano\meter} \times \SI[separate-uncertainty]{800(17)}{\nano\meter}$ (FWHM) in horizontal and vertical directions, respectively.
The theoretically expected size is $d = \SI{330}{\nano\meter}$.
The focal points in both directions do not coincide, and the lenses show a pronounced astigmatism.
\begin{figure}[!htbp]
\centering
\includegraphics[width=10cm]{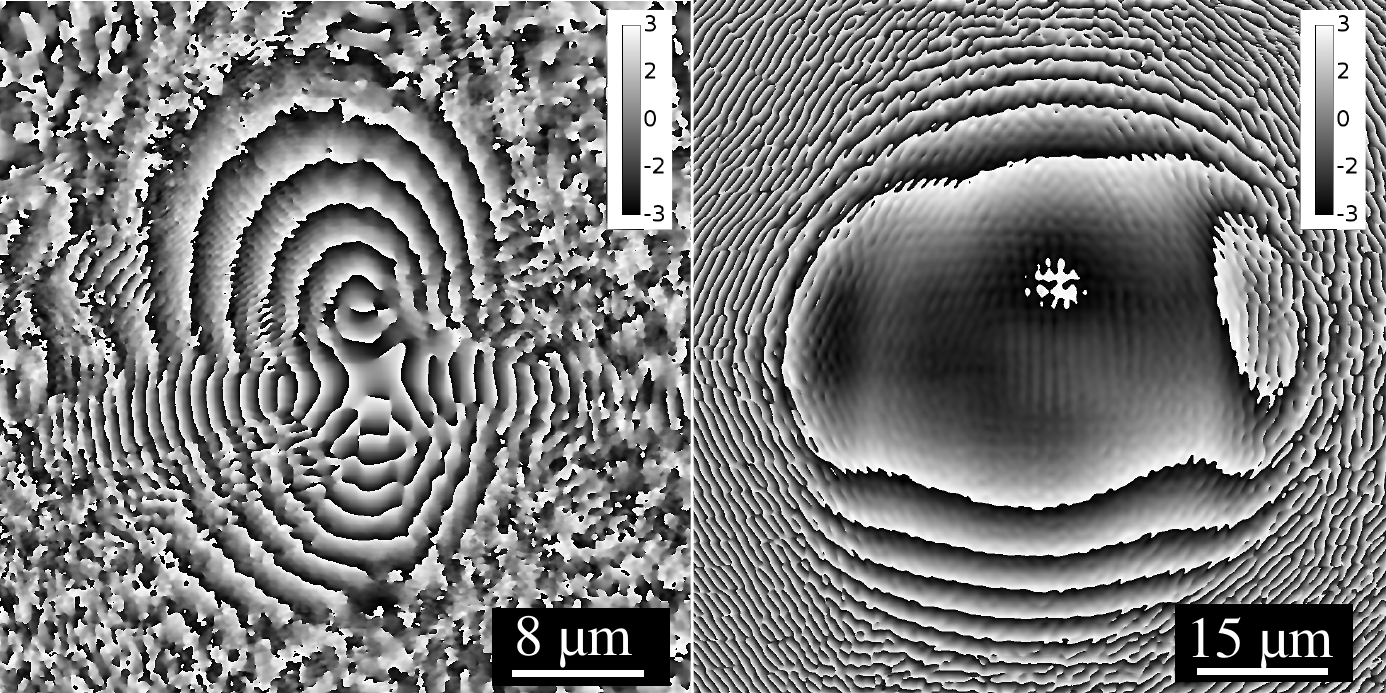}
\caption{Left: Phase profile of the beam at the focal plane. Right: Phase error induced by horizontally organized $9$-CRL (phase profile of the beam with substracted spherical wave)}
\label{9lens}
\end{figure}
Figures \ref{9lens_prof} (c) and (d) show the beam profiles in horizontal and vertical directions, respectively, and their cross sections, for the $8$-CRL.
The size of the focal spot (FWHM) is $d = \SI[separate-uncertainty]{640(20)}{\nano\meter} \times \SI[separate-uncertainty]{880(20)}{\nano\meter}$ in horizontal and vertical directions, respectively.
The theoretically expected size is $d = \SI{350}{\nano\meter}$.
The beam profile shows the same type of aberrations as $9$-CRL.
The mismatching of focus sizes and focal positions and the presence of aberrations of both lenses are most likely due to errors of the manufacturing process.
As can be seen in \Fig{MSU_SEM}, the aperture of the lenses is stretched in the vertical direction, explaining the longer focal length in this direction.
This assumption is in full agreement with the phase profile of both the focal spot and the lens exit plane, see \Fig{9lens_prof} -- the beam profile has an elliptical shape (astigmatism).
\FloatBarrier

\section{Conclusions}

We have presented a comprehensive characterization of two sets of plastic lenses manufactured via two-photon polymerization lithography.
The two sets of lenses are made of two different types of materials: IP-S and ORMOCOMP.
The first lens set, with the optical axis perpendicular to the substrate, demonstrated a very good performance and the almost absence of spherical aberrations.
Nevertheless, the increasing number of lenses in a row will result in the introduction of spherical aberrations due to shrinking of the material during printing.
Shrinking of the polymer structure occurs during the developing step due to evaporation of the solvent.
In order to keep the performance of the lens set at the highest level, the design of the parabolic shape of the lens has to be corrected in order to compensate for this.

The second set of lenses, with the optical axis parallel to the substrate suffered from deviations from an accurate parabolic profile.
Strong astigmatism coupled with spherical aberrations significantly limited the performance of these lenses.
The mediocre beam quality may be explained by the observable defects and the elliptical shape of the entrance aperture.
The latter may be explained by the displacement of the specimen along the axis of the printing beam during the exposure.
Additional deformations could have been induced by shrinking of the polymer material.
However the changes in the dimensions are not isotropic.
The lower layers of the printed structures cannot shrink freely, because they are attached to the substrate.
Thus, shrinking of the material leads to anisotropic deformation of the CRLs.
Due to the fact that the deformations are present in each individual lens, the aberrations are accumulating with the number of lenses.
As a result, in the case of horizontal orientation, the manufacturing process becomes more demanding and sensitive to printing parameters.
In order to improve the performance of this variant of CRL design the printing process should be accurately corrected.

\section*{Acknowledgments}

Authors are very grateful to Anatoly Snigirev for very fruitful discussions.
We thank Diamond Light Source for access to beamline I13-1 (MT17460-2) that contributed to the results presented here.
We acknowledge DESY (Hamburg, Germany), a member of the Helmholtz Association HGF, for the provision of experimental facilities.
Parts of this research were carried out at PETRA III and we would like to thank Gerald Falkenberg and Jan Garrevoet for assistance in using P06 beamline.

\section*{Funding}
This work was partially supported by Russian
Ministry of Education and Science (Nº 14.W03.31.0008, horizontal train lens manufacturing), Russian Science Foundation (Nº 15-12-00065, scanning electron microscopy, horizontal train lens manufacturing) and MSU Quantum Technology Center.
Frank Seiboth acknowledges funding by the Volskwagen Foundation.


\begin{thebibliography}{10}
\newcommand{\enquote}[1]{``#1''}

\bibitem{SKSL1996}
A.~Snigirev, V.~Kohn, I.~Snigireva, and B.~Lengeler, \enquote{{A compound
  refractive lens for focusing high energy x-rays},} Nature
  \textbf{384}, 49 (1996).

\bibitem{LS2016}
M.~Lyubomirskiy and C.~G. Schroer, \enquote{{Refractive Lenses for Microscopy
  and Nanoanalysis},} Synchr. Rad. News \textbf{29}, 21--26 (2016).

\bibitem{SKLGKBRSSS2002}
C.~G. Schroer, M.~Kuhlmann, B.~Lengeler, T.~F. G{\"u}nzler, O.~Kurapova,
  B.~Benner, C.~Rau, A.~S. Simionovici, A.~Snigirev, and I.~Snigireva,
  \enquote{{Beryllium parabolic refractive x-ray lenses},}
  Proc. SPIE \textbf{4783}, 10--18 (2002).

\bibitem{SHPSU2013}
A.~Schropp, R.~Hoppe, J.~Patommel, F.~Seiboth, F.~Uhl{\'e}n, U.~Vogt, H.~J.
  Lee, B.~Nagler, E.~C. Galtier, U.~Zastrau, B.~Arnold, P.~Heimann, J.~B.
  Hastings, and C.~G. Schroer, \enquote{{Scanning coherent x-ray microscopy as
  a tool for XFEL nanobeam characterization},} Proc. SPIE \textbf{8849}, 
  88490R-1 -- 88490R-08 (2013).

\bibitem{SSSWRWUNRPGVWRBGFGLNS2017}
F.~Seiboth, A.~Schropp, M.~Scholz, F.~Wittwer, C.~R{\"o}del, M.~W{\"u}nsche,
  T.~Ullsperger, S.~Nolte, J.~Rahom{\"a}ki, K.~Parfeniukas, S.~Giakoumidis,
  U.~Vogt, U.~Wagner, C.~Rau, U.~Boesenberg, J.~Garrevoet, G.~Falkenberg, E.~C.
  Galtier, H.~J. Lee, B.~Nagler, and C.~G. Schroer, \enquote{{Perfect {X-ray}
  focusing via fitting corrective glasses to aberrated optics},} Nature
  Commun. \textbf{8}, 14623 (2017).

\bibitem{SKPBFLBRVHK2006}
C.~G. Schroer, O.~Kurapova, J.~Patommel, P.~Boye, J.~Feldkamp, B.~Lengeler,
  M.~Burghammer, C.~Riekel, L.~Vincze, A.~{van der Hart}, and M.~K{\"u}chler,
  \enquote{{Hard X-Ray Nanoprobe based on Refractive X-Ray Lenses},} AIP Conf. Proc. \textbf{879}, 1295--1298 (2006).

\bibitem{SBFPSSSSFWR2009}
C.~G. Schroer, P.~Boye, J.~M. Feldkamp, J.~Patommel, D.~Samberg, A.~Schropp,
  A.~Schwab, S.~Stephan, G.~Falkenberg, G.~Wellenreuther, and N.~Reimers,
  \enquote{{Hard {X}-ray Nanoprobe at Beamline {P06} at {PETRA III}},} Nucl.
  Instrum. Meth. A \textbf{616}, 93--97 (2010).

\bibitem{SSLJHDP2016}
H.~Simons, F.~St{\"o}hr, J.~Michael-Lindhard, F.~Jensen, O.~Hansen, C.~Detlefs,
  and H.~F. Poulsen, \enquote{{Full-field hard x-ray microscopy with
  interdigitated silicon lenses},} Opt. Commun. \textbf{359}, 460--464
  (2016).

\bibitem{SL2005c}
C.~G. Schroer and B.~Lengeler, \enquote{{Focusing Hard X Rays to Nanometer
  Dimensions by Adiabatically Focusing Lenses},} Phys. Rev. Lett. \textbf{94},
  054802 (2005).

\bibitem{PKHRSWJRWBSSBFS2017}
J.~Patommel, S.~Klare, R.~Hoppe, S.~Ritter, D.~Samberg, F.~Wittwer, A.~Jahn,
  K.~Richter, C.~Wenzel, J.~W. Bartha, M.~Scholz, F.~Seiboth, U.~Boesenberg,
  G.~Falkenberg, and C.~G. Schroer, \enquote{{Focusing Hard X Rays Beyond the
  Critical Angle of Total Reflection by Adiabatically Focusing Lenses},} Appl.
  Phys. Lett. \textbf{110}, 101103 (2017).

\bibitem{PBAKSBEKLYPSFS2017}
A.~K. Petrov, V.~O. Bessonov, K.~A. Abrashitova, N.~G. Kokareva, K.~R.
  Safronov, A.~A. Barannikov, P.~A. Ershov, N.~B. Klimova, I.~I. Lyatun, V.~A.
  Yunkin, M.~Polikarpov, I.~Snigireva, A.~A. Fedyanin, and A.~Snigirev,
  \enquote{{Polymer X-ray refractive nano-lenses fabricated by additive
  technology},} Opt. Express \textbf{25}, 14173--14181 (2017).

\bibitem{app8050737}
T.~dos Santos~Rolo, S.~Reich, D.~Karpov, S.~Gasilov, D.~Kunka, E.~Fohtung,
  T.~Baumbach, and A.~Plech, \enquote{{A Shack-Hartmann Sensor for Single-Shot
  Multi-Contrast Imaging with Hard X-rays},} Appl. Sci. \textbf{8}
  (2018).

\bibitem{FW2013}
J.~Fischer and M.~Wegener, \enquote{{Three-dimensional optical laser
  lithography beyond the diffraction limit},} Laser Photonics Rev.
  \textbf{7}, 22--44 (2013).

\bibitem{SBFHPSSGWSGMVWSBS2010}
A.~Schropp, P.~Boye, J.~M. Feldkamp, R.~Hoppe, J.~Patommel, D.~Samberg,
  S.~Stephan, K.~Giewekemeyer, R.~N. Wilke, T.~Salditt, J.~Gulden, A.~P.
  Mancuso, I.~A. Vartanyants, E.~Weckert, S.~Sch{\"o}der, M.~Burghammer, and
  C.~G. Schroer, \enquote{{Hard x-ray nanobeam characterization by coherent
  diffraction microscopy},} Appl. Phys. Lett. \textbf{96}, 091102 (2010).

\bibitem{KTDBMVJP2010}
C.~M. Kewish, P.~Thibault, M.~Dierolf, O.~Bunk, A.~Menzel, J.~Vila-Comamala,
  K.~Jefimovs, and F.~Pfeiffer, \enquote{{Ptychographic characterization of the
  wavefield in the focus of reflective hard {X}-ray optics},} Ultramicroscopy
  \textbf{110}, 325--329 (2010).

\bibitem{KGLQSBKVBFMA2010}
C.~M. Kewish, M.~Guizar-Sicairos, C.~Liu, J.~Qian, B.~Shi, C.~Benson, A.~M.
  Khounsary, J.~Vila-Comamala, O.~Bunk, J.~R. Fienup, A.~T. Macrander, and
  L.~Assoufid, \enquote{{Reconstruction of an astigmatic hard X-ray beam
  alignment of {K-B} mirrors from ptychographic coherent diffraction data},}
  Opt. Express \textbf{18}, 23420--23427 (2010).

\bibitem{HHPSSSBS2011c}
S.~H{\"o}nig, R.~Hoppe, J.~Patommel, A.~Schropp, S.~Stephan, S.~Sch{\"o}der,
  M.~Burghammer, and C.~G. Schroer, \enquote{{Full optical characterization of
  coherent x-ray nanobeams by ptychographic imaging},} Opt. Express
  \textbf{19}, 16325--16329 (2011).

\bibitem{VDGMKMBD2011}
J.~Vila-Comamala, A.~Diaz, M.~Guizar-Sicairos, A.~Mantion, C.~M. Kewish,
  A.~Menzel, O.~Bunk, and C.~David, \enquote{{Characterization of
  high-resolution diffractive X-ray optics by ptychographic coherent
  diffractive imaging},} Opt. Express \textbf{19}, 21333--21344 (2011).

\bibitem{KBNKPHLS2014}
A.~Kubec, S.~Braun, S.~Niese, P.~Kr{\"u}ger, J.~Patommel, M.~Hecker, A.~Leson,
  and C.~G. Schroer, \enquote{{Ptychography with mulitlayer {Laue} lenses},} J.
  Synchrotron Rad. \textbf{21}, 1122--1127 (2014).

\bibitem{SSPHWRSKJRBFS2014}
F.~Seiboth, M.~Scholz, J.~Patommel, R.~Hoppe, F.~Wittwer, J.~Reinhardt,
  J.~Seidel, M.~Knaut, A.~Jahn, K.~Richter, J.~W. Bartha, G.~Falkenberg, and
  C.~G. Schroer, \enquote{{Hard x-ray nanofocusing by refractive lenses of
  constant thickness},} Appl. Phys. Lett. \textbf{105}, 131110 (2014).

\bibitem{MPAKMPGBBBOYACB2015}
A.~J. Morgan, M.~Prasciolu, A.~Andrejczuk, J.~Krzywinski, A.~Meents,
  D.~Pennicard, H.~Graafsma, A.~Barty, R.~J. Bean, M.~Barthelmess,
  D.~Oberthuer, O.~Yefanov, A.~Aquila, H.~N. Chapman, and S.~Bajt,
  \enquote{{High numerical aperture multilayer {Laue} lenses},} Sci. Rep. \textbf{5}, 09892 (2015).

\bibitem{LSBGGKMZ2002c}
B.~Lengeler, C.~G. Schroer, B.~Benner, A.~Gerhardus, T.~F. G{\"u}nzler,
  M.~Kuhlmann, J.~Meyer, and C.~Zimprich, \enquote{{Parabolic refractive
  {X-ray} lenses},} J. Synchrotron Rad. \textbf{9}, 119--124 (2002).

\end{thebibliography}
\end{document}